# Towards the Framework of Information Security

**Dr. Brijender Kahanwal[1*] and Dr. Tejinder Pal Singh[2]**

[1*]Assistant Professor, CSE Department, Faculty of Engineerig, Galaxy Global Group of Institutions, Dinarpur, Sahabad, Ambala, Haryana, India
 imkahanwal@gmail.com

[2]HOD and Assistant Professor, Applied Sciences Department, R. P. Inderprasth Institute of Technology (RPIIT), Bastara, Karnal, Haryana, India
 tps5675@gmail.com

*Abstract*

Today's modern society is extremely dependent on computer based information systems. Many of the organizations would simply not be able to function properly without services provided by these systems, just like financing organizations. Although interruption might decrease the efficiency of an organization, theft or unintentional disclosure of entrusted private data could have more serious consequences, such as legal actions as well as loss of business due to lack of trust from potential users. This dependence on information systems has lead to a need for securing these systems and this in turn has created a need for knowing how secure they are.

The introduction of the information society has changed how people interact with government agencies. Government agencies are now encouraged to uphold a 24-hour electronic service to the citizens. The introduction of government services on the Internet is meant to facilitate communication with agencies, decrease service times and to lessen the amount of papers that needs to be processed. The increased connectivity to the Internet results in a rising demand for information security in these systems. In this article, we have discussed about many file data breaches in the past and current history and they are going to increase day by day as the reports by DataLossDB [1] (Open Security Foundation) organization, a non-profit organization in US.

## INTRODUCTION

The modern society is totally relying on the information. The complete business revolves around it. The business's success or failure depends upon it. If a business man is aware about the importance of the information then there is less number of chances to fail. In this article we have gone through the consequences of the information security if that is compromised by the society. The DataLossDB organization is giving good insight look at the details of losses in the complete world. This is a non-profit organization of the United States. In this article first of all we will be introduced by the definitions of information, information security. Also we will know the importance of information security and the consequences of information security, if we will compromise it. The DataLossDB (http://datalossdb.org/) [1] is a US research project (but also records global data losses which include the UK and Europe) aimed at documenting known and reported data loss incidents world-wide. The effort is now a community one, and with the move to Open Security Foundation's DataLossDB.org, asks for contributions of new incidents and new data for existing incidents. This is a very useful resource for anyone wanting to find out about the latest breaches.

## INFORMATION AND ITS CLASSIFICATION

Information comprises the meanings and interpretations that people place upon facts, or data. The value of information springs from the ways it is interpreted and applied to make products, to provide services, and so on.

Many modern writers look at organizations in terms of the use they make of information. For instance, one particularly successful model of business is based on the assets that a firm owns. Assets have traditionally meant tangible things like money, property, plant, systems; but business analysts have increasingly recognized that information is itself an asset, crucial to adding value. But, of course, there is a negative side too: the use of information in both the for-profit and not-for- profit sectors is increasingly the subject of legislation and regulation, in recognition of the damage its misuse can have on individuals.

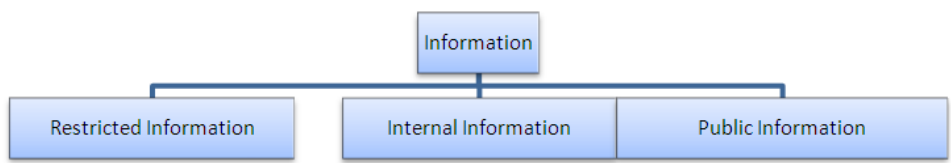

Figure 1.1: Classification of the information

Information classification is an understanding of what we are going to protect. It may be classified in the following three categories and also shown in the Figure 1.1:

*Restricted information:* Personal information is the restricted one e. g. social security number, driver's license, financial account numbers etc.

*Internal information:* It is the information category which is accessible only with the reason to know it, e. g. unpublished research paper, employees benefit statements, students financial aid statements.

*Public information:* This type of information is related to the general public, e. g. any Cricket / Football games schedules, examination schedule etc.

**INFORMATION SECURITY**

‒*Information security* is all about protecting and preserving information. It's all about protecting and preserving the confidentiality, integrity, authenticity, availability, and reliability of information is given by ISO 27001 [7].

Seen in the way we have just defined it, information is a valuable asset. Information security protects information (and the facilities and systems that store, use and transmit it) from a wide range of threats, in order to preserve its value to an organization.

This definition of information security is adapted from that of the American National Security Telecommunications and Information Systems Security Committee (NSTISSC) [9]. There are two important characteristics of information that determine its value to an organization:

- Scarcity of the information outside the organization;
- Shareability of the information within the organization, or some part of it.

Simplifying somewhat, these characteristics state that information is only valuable if it provides advantage or utility to those who have it, compared with those who don't. Thus the value of any piece of information relates to its levels of shareability and scarcity. The aim of information security is to preserve the value of information by ensuring that these levels are correctly identified and preserved.

‒Information security means protecting information and information systems from unauthorized access, use, disclosure, disruption, modification, perusal, inspection, recording or destruction.‖

Threats to information influence the organization's ability to share it within, or to preserve its scarcity outside. And threats that are carried out can cost millions in compensation and reputation, and may even jeopardize an institution's ability to survive. Here are some examples in which the making available of information that should have been kept scarce or the restricting of information that should have been shareable has damaged an organization.

When we leave our house for work in the morning, we probably take steps to protect it and the contents from unauthorized access, damage and theft (e.g. turning off the lights, locking the doors and setting the alarm). This same principle can be applied to information – steps must be put in place to protect it. If left unprotected, information can be accessed by anyone. If information should fall into the wrong hands, it can wreck lives, bring down businesses and even be used to commit harm. Quite often, ensuring that information is appropriately protected is both a business and legal requirement. In addition, taking steps to protect your own personal information is a matter of privacy retention and will help prevent identity theft.

**IMPORTANCE OF INFORMATION SECURITY**

Information security is the collection of technologies, standards, policies and management practices that are applied to information to keep it secure.

But why is it important to secure information? And how should its security is managed? To start thinking about these questions, consider the following statements about information:

Some typical information is held by both businesses and individuals. At the very least, businesses will hold sensitive information on their employees, salary information, financial results, and business plans for the year ahead. They may also hold trade secrets, research and other information that gives them a competitive edge. Individuals usually hold sensitive personal information on their home computers and typically perform online functions such as banking, shopping and social networking; sharing their sensitive information with others over the internet.

As more and more of this information is stored and processed electronically and transmitted across company networks or the internet, the risk of unauthorized access increases and we are presented with growing challenges of how best to protect it.

In today's high technology environment, organizations are becoming more and more dependent on their information systems. The public is increasingly concerned about the proper use of information, particularly personal data. The threats to information systems from criminals and terrorists are increasing. Many organizations will identify information as an area of their operation that needs to be protected as part of their system of internal control.

Information security is the process of preventing and detecting unauthorized use of your important data. Prevention measures help you to stop unauthorized users (also known as "intruders") from accessing any part of your computer data.

We use computers for everything from banking and investing to shopping and communicating with others through email or chat programs. Although you may not consider your communications "top secret," you probably do not want strangers reading your email, using your computer to attack other systems, sending forged email from your computer, or examining personal information stored on your computer (such as financial statements).

Intruders (also referred to as hackers, attackers, or crackers) may not care about your identity. Often they want to gain control of your computer so they can use it to launch attacks on other computer systems.

It is vital to be worried about information security because much of the value of a business is concentrated in the value of its information. Information security affects every structural and behavioral aspect of an organization. Each individual that interacts with an organization in any way – from the potential customer browsing the website, to the managing director; from the malicious hacker, to the information security manager – will make his or her own positive or negative contribution to the information security of the organization.

**THE CONSEQUENCES OF INFORMATION BEING COMPROMISED**

When information is not adequately protected, it may be compromised and this is known as an information or security breach. The consequences of an information breach are severe. For businesses, a breach usually entails huge financial penalties, expensive law suits, loss of reputation and business. For individuals, a breach can lead to identity theft and damage to financial history or credit rating. Recovering from information breaches can take years and the costs are huge. According to the Ponemon institute, the average cost of an information breach during 2008 was $202 per record breached. So, if 100,000 records were breached, the average cost for this breach would be $20 million! 70% of this cost is down to lost business as a result of the breach.

Data security has become a very important issue with growing dependence on storage systems and increasing reliance on the internet for communication. Millions of dollars have been reported to be lost due to security breaches. The key factors of interest for a deployment of a storage solution are security, performance and usability.

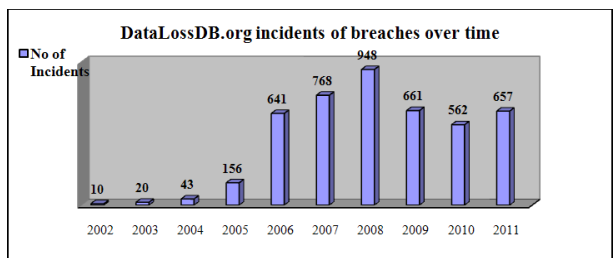

Figure 1.2: DataLossDB.org breach incidents over time till September 2011.

In the above Figure 1.2 all the breach incidents are shown from the year 2002 to September 2011. In the starting year of study they were very less only 10 incidents in the complete year which are to be considered by the organization. And after that they start to increase till 2008, the year of highest number of incidents takes place. There is slight down fall till the year 2010. But again they start to increase in the current year 2011. The results are till the month of September, there may be more one till the end of the year.

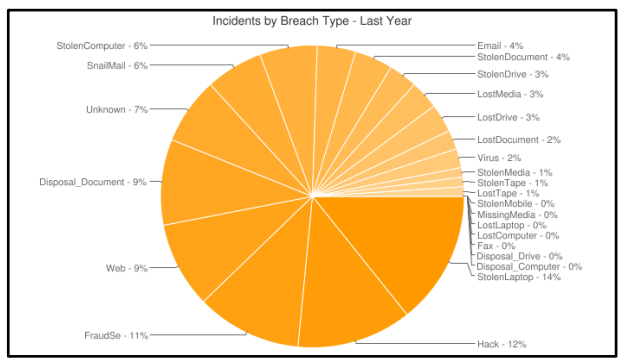

Figure 1.3: DataLossDB.org breach incidents by Breach type in the last year 2010.

The Figure 1.3 shows the breach types in the incidents in the last year 2010. In this pie chart 29% of incidents takes place by the stolen of the hardware (computer=6% + laptop=14% + document (hardcopy) =4% + drive=3% + media=1% + tape=1% + mobile=0%). It is the largest area. Second largest part is occupied by the hacking that is 12%. The third largest is the fraudSe which is 11% in total. Next largest is web and disposal document that is 9% each. By the snail mail is 6% of total. By the email is 4% loss. Unknown activities also take a share that is 7% of total. The loss also take place when the users forgets the information that shares 9% of total incidents (media=3% + drive=3% + document=2% + tape=1%). Virus shares only 2% part of the total incidents. In conclusion of the above pie chart we must protect our hardware from the theft.

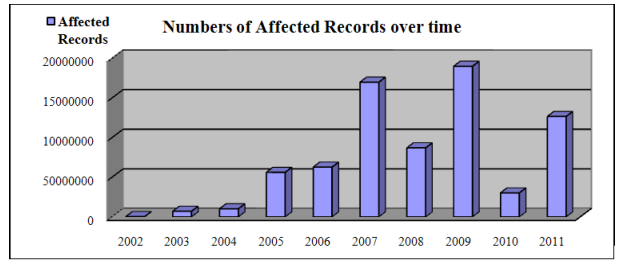

Figure 1.4: DataLossDB.org affected records by the breaches over the time.

In the Figure 1.4 the continuous increase is from the year 2002 to the year 2007 and the most of the records were affected in the year 2009. In the year 2010 there is a great downfall in the number of affected records. But they are going to increase in the year 2011 till August month.

Table 1.1: DataLossDB.org affected records by breaching over time.

| YEAR | Total Records Affected | YEAR | Total Records Affected |
|---|---|---|---|
| 2002 | 321,824 | 2007 | 170,020,143 |
| 2003 | 7,061,985 | 2008 | 87,106,008 |
| 2004 | 10,199,081 | 2009 | 190,099,668 |
| 2005 | 56,202,442 | 2010 | 30,247,027 |
| 2006 | 62,764,205 | 2011 | 126,714,659 |

In conclusion you may think that the likelihood of an information breach affecting you is rare, but this could not be further from the truth. Just take a look at the link http://www.privacyrights.org/data-breach that lists recently recorded information breaches in the US. The section highlighted the major areas where we must take care for the information security.

**CONCLUSION**

With the help of the above article we have highlighted the main information breaches in the past and the current one. The society must be aware about the losses which are going to increase day by day. We must develop the information systems which are being developed to apply all the security principles. Watching out such types of breaches I have developed the Java File Security System (JFSS) [3][4][5][6] that is an encryption file system. To get more knowledge about it you may follow the references.